\documentclass[twocolumn,preprintnumbers,amsmath,amssymb]{revtex4}

\usepackage{epsfig}
\usepackage{subfigure}
\usepackage{graphics}
\usepackage{amsmath}
\usepackage{bbm}

\usepackage{varioref}
\usepackage{float}
\usepackage{units}
\usepackage{textcomp}
\usepackage{amsmath}
\usepackage{amssymb}
\usepackage{graphicx}
\usepackage{setspace}
\usepackage{esint}
\usepackage{feyn}

\begin{document}

\title{Microcausality in strongly interacting  fields}
\author{L.~Rauber}%
\affiliation{%
 Institut f{\"u}r Theoretische Physik, %
  Universit\"at Giessen, %
  35392 Giessen, %
  Germany %
}

\author{W.~Cassing}
\email{Wolfgang.Cassing@theo.physik.uni-giessen.de}
\affiliation{%
  Institut f{\"u}r Theoretische Physik, %
  Universit\"at Giessen, %
  35392 Giessen, %
  Germany %
}
\date{\today}
\begin{abstract}
We study the properties of strongly interacting massive quantum
fields in space-time as resulting from a parametric decay of the
fields with a large decay width $\gamma$. The resulting imaginary
part of the retarded and advanced propagators in this case is of
Lorentzian form and the theory conserves microcausality, i.e.
the commutator between the fields vanishes
for space-like distances in space-time. However, when considering
separately space-like and time-like components of the spectral
function in momentum space we find microcausality to be violated for
each component separately. This implies that the modeling of
effective field theories for strongly interacting systems has to be
considered with great care and restrictions to time-like four
momenta in case of broad spectral functions have to be ruled out.
Furthermore, when employing effective propagators with a width
$\gamma({\bf p}^2)$
depending explicitly on three-momentum ${\bf p}$ the commutator of
the fields no longer vanishes for $r>t$ since the related field
theory becomes nonlocal and violates microcausality.
 \\ \\

\end{abstract}
\maketitle
\section{Introduction}
Quantum Chromo-Dynamics (QCD) is considered to be the theory of the
strong interaction, however, is accessable to perturbation theory
only in the limit of short distances or high momentum transfer,
respectively. Thermodynamical properties of hadronic or partonic
matter at finite temperature $T$ and/or chemical potential $\mu_q$
involve large distance interactions and can rigorously only be
addressed by lattice QCD (predominantly at vanishing chemical
potential) in Euclidean space. Alternatively, one might employ
effective field theories that share the symmetry properties of QCD
and fix the couplings to reproduce field expectation values and
correlators \cite{ef0,ef1,ef2,ef3,ef4,ef5,ef6,ef7,ef8,ef9,ef10}. In
fact, the knowledge about the phase diagram of strongly interacting
hadronic/partonic matter has been increased substantially in the
last decades. At vanishing (or low) chemical potentials lattice QCD
(lQCD) calculations have provided reliable results  on the equation
of state \cite{lQCD,Peter1} and given a glance at the transport
properties (or correlators
\cite{l0,ll0,l1,l2,l3,l4,l5,Aarts,lqcdb0,lqcdb1,lqcdb2,lqcdb3}) in
particular in the partonic phase.

Recent studies of 'QCD matter' in equilibrium -- using lattice QCD
calculations \cite{l0,ll0} or partonic transport models in a finite
box with periodic boundary conditions \cite{Vitalii1} -- have
demonstrated that the ratio of the shear viscosity to entropy
density $\eta/s$ should have a minimum close to the critical
temperature $T_c$, similar to atomic and molecular systems
\cite{review}. On the other hand, the ratio of the bulk viscosity to
the entropy density $\zeta/s$ should have a maximum close to $T_c$
\cite{Vitalii1} or might even diverge at $T_c$
\cite{MaxBulk1,MaxBulk2,MaxBulk3,MaxBulk4,MaxBulk5}. Indeed, the
minimum of $\eta/s$ at $T_c \approx$ 160 MeV is close to the lower
bound of a perfect fluid with $\eta/s= 1/(4\pi)$~\cite{KSS} for
infinitely coupled supersymmetric Yang-Mills gauge theory (based on
the AdS/CFT duality conjecture). This suggests the `hot QCD matter'
to be the `most perfect fluid'~\cite{new1,new2,Barbara}. On the
other hand the transport studies in Refs.
\cite{Vitalii1,Ca13,Steinert} have provided results for the shear
and bulk viscosity as well as the electric conductivity that are
very close to lattice QCD results, however, employ the notion of a
strongly interacting gas of quasiparticles with a dynamically
generated mass that is sufficiently larger than the width of their
spectral functions. These studies have been based on the Dynamical
QuasiParticle Model (DQPM) \cite{Peshier,DQPM1} that incorporates
effective propagators for the partons  with a finite width of the
spectral functions $A_i (\omega_i, {\bf p}_i)$, i.e. for scalar
fields (${\tilde p} = (\omega,
{\bf p})$)
$$A_i (\omega_i, {\bf p}_i)  \ = \frac{\gamma_i}{2\tilde{E}_i}
\biggl(\frac{1}{(\omega_i - \tilde{E}_i)^2 + \gamma_i^2} -
\frac{1}{(\omega_i + \tilde{E}_i)^2 + \gamma_i^2} \biggr)  $$
\begin{equation}
\label{eq1}
 = \frac{2 \omega_i \gamma_i}{(\omega_i^2
- {\bf p}_i^2 - M_i^2)^2 +
4 \gamma_i^2 \omega_i^2}, \end{equation}
 with $\tilde{E}_i^2 ({\bf p}_i) = {\bf p}_i^2 + M_i^2 - \gamma_i^2$ and $i \in [g, q, \bar{q}]$.
 The spectral functions $A_i (\omega_i)$ are antisymmetric in $\omega_i$ and normalized as:
\begin{equation}
\label{equ:Sec2.3} \int_{- \infty}^{+ \infty} \frac{d \omega_i}{2 \pi} \ 2\omega_i \ A_i (\omega_i, {\bf p})
= 1 ,
\end{equation}
where  $M_i$ and $\gamma_i$ are the dynamical quasiparticle mass (i.e.
pole mass) and width of the spectral function for particle $i$,
respectively. They are directly related to the real and imaginary
parts of the related self-energy, e.g. $\Pi_i = M_i^2 - 2 i \gamma_i
\omega_i$, \cite{DQPM1}. In the off-shell approach, $\omega_i$ is an
independent variable { and} related to the \emph{``running mass''}
$m_i$ by: $\omega_i^2 = m_i^2 + {\bf p}_i^2$. In case of vector
fields or fermion fields the following retarded propagators are
employed that differ from the 'free' massive case only by the
additional $(2  \gamma_V \omega)^2$ or $\pm i \gamma_F$ in the
denominator and corresponding matrices in the numerator
\cite{Peter2}:
\begin{equation} \label{eq2}
A_V^{\mu \nu}(\omega, {\bf p},\gamma_V)= \gamma_V \frac{2 \omega (g^{\mu \nu} - p^\mu p^\nu/M_V^2)}{(\omega^2
- {\bf p}^2 - M_V^2)^2 + 4 \gamma_V^2 \omega^2},
\end{equation}
and
\begin{eqnarray}
&& A_F(\omega, {\bf p},\gamma_F) = \frac{1}{4 E_F}  \label{eq3} \\
&& \times \left( \frac{E_F\gamma^0 + {\bf p} \cdot {\mathbf \gamma} +
m_F}{\omega-E_F - i \gamma_F } -\frac{-E_F\gamma^0 + {\bf p} \cdot
{\mathbf \gamma} + m_F}{\omega+E_F - i \gamma_F } \right. \nonumber\\
&& \left. -\frac{E_F\gamma^0 + {\bf p} \cdot {\mathbf \gamma} +
m_F}{\omega-E_F + i \gamma_F } + \frac{-E_F\gamma^0 + {\bf p} \cdot
{\mathbf \gamma} + m_F}{\omega+E_F + i \gamma_F } \right)
\nonumber
\end{eqnarray}
 with $E_F= {\bf p}^2+m_F^2$ in obvious notation.

As is seen e.g. from the spectral function (\ref{eq1}) it is
non-vanishing for time-like (${\tilde p}^2 >0$) as well as for space-like
(${\tilde p}^2< 0$) four-momenta such that the question emerges if the
theoretical concept behind the DQPM  (or other effective approaches)
conserves microcausality, i.e. that the
spectral function transformed to space-time has only support on
and within the lightcone. We recall that the Fourier transform of the
spectral function $A(\omega, {\bf p})$ is proportional to the
commutator of the fields at different space-time point and its
integration over energy $\omega$ ensures a proper quantization (see below).
This is of particular importance since a
transport realization can only propagate 'quasiparticles' within
or on the lightcone \cite{PHSD1,PHSD2}. More importantly, in the DQPM spectral
contributions are separated into  time-like (${\tilde p}^2 >0$) and space-like
(${\tilde p}^2< 0$) four-momentum parts and the additional question arises
if the contributions separately conserve microcausality.

The layout of our study is as follows: In Section II we briefly
present the basic definitions and relations between retarded and
advanced propagators and recall the analytic proof for
microcausality in case of the spectral functions
(\ref{eq1}),(\ref{eq2}),(\ref{eq3}). In Section III the actual
problem is set up for time-like (${\tilde p}^2 >0$) and space-like
(${\tilde p}^2< 0$) four-momentum parts of the spectral function and
its numerical realization. Furthermore, we present the actual
numerical results for strong coupling and investigate the aperiodic
limit as well as the case $\gamma > M$. A summary and discussion of
results is given in Section IV.

\section{Propagators and spectral functions}
In this work we will concentrate on the model case of a massive scalar
field coupled e.g. to an external fermion field
( $\sim \partial_\mu \Phi({x}) {\bar \Psi}({x})
\gamma^\mu \Psi({x})$ with a vanishing three-current, i.e.
the field equation
\begin{equation}
\left(\frac{\partial^{2}}{\partial t^{2}}-\bigtriangleup+M^{2}
+2\gamma\frac{\partial}{\partial
t}\right)\Phi({x})=0,
\label{eq:KleinGordon}
\end{equation}
where  $\gamma$ stands for the strength of the coupling (e.g. $g_s <\Psi^\dagger \Psi>/2$).
Eq. (\ref{eq:KleinGordon}) has the algebraic solution
\begin{equation}
\widetilde{G}(\mathbf{p})=\frac{-1}{\omega^{2}-\boldsymbol{p}^{2}-M^{2}+2i\gamma\omega}
,
\label{eq:G(p)AlgebraischGewonnen}
\end{equation}
which leads to the retarded Green-function
 $G_{\mathrm{ret}}$ obeying
\begin{equation}
G_{\mathrm{ret}}({x}-{y})=0\ \mathrm{for}\
x^{0}-y^{0}<0
\label{eq:retardierteRandbedingung}
\end{equation}
by a 4-dimensional Fourier transformation of
(\ref{eq:G(p)AlgebraischGewonnen}),
\begin{equation} \label{eq5}
G_{\mathrm{ret}}({x})= \int \frac{d^4 {\tilde p}}{(2 \pi)^4} \
\widetilde{G}({\tilde p})\ \exp(-i {\tilde p} x) .
\end{equation}
We point out that $\Im {\tilde G}({\tilde p})$ is identical to (\ref{eq1}).
We recall, furthermore, that solutions of the Kadanoff-Baym equations \cite{KB}
for $\Phi^4$- theory in 2+1 dimensions \cite{Juchem} have lead to
spectral functions that are very close to (\ref{eq1}) also for
strong coupling.

\subsection{Analytical results}
The integration over $d\omega = dp^0$ in (\ref{eq5}) can be carried out by contour integration
and the angular integration in three-momentum is straight forward.
With $\mu=\sqrt{M^{2}-\gamma^{2}}$ the remaining integral kernel
reads ($p=|{\bf p}|$)
\begin{align}
K(x):= &
\frac{1}{\left|\boldsymbol{x}\right|}\intop_{0}^{\infty} p
\frac{\sin(t\sqrt{\mu^{2}+p^{2}})}{\sqrt{\mu^{2}+p^{2}}} \sin(\left|\boldsymbol{x}\right|p)\,\mathrm{d}p\\
= & \frac{1}{2\left|\boldsymbol{x}\right|}\intop_{-\infty}^{\infty}
p
\frac{\sin(t\sqrt{\mu^{2}+p^{2}})}{\sqrt{\mu^{2}+p^{2}}}\sin(\left|\boldsymbol{x}\right|p)\,\mathrm{d}p\nonumber
,
\end{align}
which has a singular contribution on the lightcone and a regular
part on and within the lightcone. The remaining integration over
$dp$ gives for the retarded {Green}-function (using
$x=(t,\boldsymbol{x})=(x^0,\boldsymbol{x})$)
\begin{equation}
G_{\mathrm{ret}}({x})=\frac{e^{-\gamma
t}\Theta(t)}{2\pi} \delta\left(t^{2}-\boldsymbol{x}^{2}\right)
\end{equation} $$ -\frac{e^{-\gamma
t}\Theta(t)}{4\pi} \Theta\left(t^{2}-\boldsymbol{x}^{2}\right)
\frac{\mu}{\sqrt{t^{2}-\boldsymbol{x}^{2}}}
J_{1}\left(\mu\sqrt{t^{2}-\boldsymbol{x}^{2}}\right) $$ for $\mu^2
\ge 0$. With
\begin{equation}
\delta\left(t^{2}-\boldsymbol{x}^{2}\right)\Theta(t)
=\delta\left((t-\left|\boldsymbol{x}\right|)(t+\left|\boldsymbol{x}\right|)\right)\Theta(t)
=
\frac{\delta(t-\left|\boldsymbol{x}\right|)}{2\left|\boldsymbol{x}\right|}
\end{equation} one arrives at the final result \cite{key-BogoliubovShirkov}
\begin{equation}
G_{\mathrm{ret}}(\boldsymbol{x})=\left(\frac{\delta\left(t-
\left|\boldsymbol{x}\right|\right)}{4\pi\left|\boldsymbol{x}\right|}-
R(t^{2}-\boldsymbol{x}^{2}) \right) e^{-\gamma t}\Theta(t)
\label{eq:TheorieErgebnis}
\end{equation}
with
\begin{equation} \label{regular1}
R(t^{2}-\boldsymbol{x}^{2})=\Theta\left(t^{2}-\boldsymbol{x}^{2}\right)\frac{\mu}{4\pi\sqrt{t^{2}-\boldsymbol{x}^{2}}}
J_{1}\left(\mu\sqrt{t^{2}-\boldsymbol{x}^{2}}\right) ,
\end{equation}
where $J_1$ is the Bessel function.
In the actual calculations the $\delta$-distribution term on the
lightcone will be subtracted and we will address the regular part
(\ref{regular1}) including the overall exponential decay in time,
i.e.
\begin{equation} \label{regular}
{\tilde R}(t^{2}-\boldsymbol{x}^{2})=R(t^{2}-\boldsymbol{x}^{2}) \
e^{-\gamma t}\Theta(t) .
\end{equation}
 We note in passing that the related results for massive vector
fields and Dirac fields read \cite{Peter}
\begin{equation} \label{GV} G_{\mathrm{ret}}^{\mu \nu}({x}) =
 \left( {g^{\mu \nu} + \frac{1}{M_V^2} \partial^\mu \partial^\nu } \right)  G_{\mathrm{ret}}({x})
\end{equation} and
\begin{equation} \label{GF}
G_{\mathrm{ret}}^F(x) =
 \left( {m_F \cdot 1_4+i \gamma^\mu \partial_\mu } \right) G_{\mathrm{ret}}({x}) .
\end{equation}
Eqs. (\ref{GV}) and (\ref{GF}) demonstrate that it is sufficient to
investigate microcausality for the scalar case since microcausality
for the scalar field implies microcausality for the corresponding
vector and fermion fields.

The retarded  {Green}-function (\ref{eq:TheorieErgebnis}) is close to the solution of the free
 massive {Klein-Gordon}-field except for the factor $e^{-\gamma t}$ describing the decay of the propagator in time
 and the reduced mass $\mu=\sqrt{M^{2}-\gamma^{2}}$ that incorporates a downward shift of the mass $M$ as
 in case of the damped harmonic oscillator.

Similar relations hold for the advanced {Green-}function which is obtained by replacing $\gamma \rightarrow - \gamma$
and a multiplication by $-1$ due to the opposite contour integration:
\begin{equation}
G_{\mathrm{av}}(x)=\left(-\frac{\delta\left(t+\left|\boldsymbol{x}\right|\right)}{4\pi\left|\boldsymbol{x}\right|}
+R\left(t^{2}-\boldsymbol{x}^{2} \right) \right) e^{+\gamma t}\ \Theta(-t)
\end{equation}
following $
G_{\mathrm{av}}({x}-{y})=0\ \mathrm{f\ddot{u}r}\
x^{0}-y^{0}>0 $. In four-momentum space the advanced propagator is
given by (\ref{eq:G(p)AlgebraischGewonnen}) replacing $\gamma$ by
$-\gamma$. Accordingly, ${\tilde G}_{ad}({\tilde p}) - {\tilde G}_{ret}({\tilde p})
= -2 i A({\tilde p})$ is purely imaginary and equal to twice the spectral
function (\ref{eq1}).

\subsection{Spectral functions}
Of central interest in our study is the scalar spectral function $A(\omega, {\bf p})$ (\ref{eq1}),
i.e. the imaginary part of the retarded propagator. The commutator between the fields at different space-time
points can also be written as the difference of advanced and retarded
propagators (due to opposite signs of the imaginary parts in the propagators \cite{xx}):
\begin{equation}
\left[\Phi({x}),\,\Phi^{\dagger}(\mathbf{0})\right]=i\Delta^{\star}({x})
=i\left(G_{\mathrm{av}}({x})-G_{\mathrm{ret}}({x})\right) =: C(x). \label{eq:ZusammenhangKommutatorGreen}
\end{equation}
Except for a factor $\exp(-\gamma t)$ the quantity $\Delta^* $ is
identical to the Schwinger $\Delta$-function $\Delta(x,\mu)$ with effective
mass $\mu$,
\begin{equation}
\Delta^{\star}({x})=\Delta({x},\,\mu)\cdot
e^{-\gamma\left|t\right|} ,
\end{equation}
\begin{equation} \label{Delta}
\Delta({x},\, \mu)=-\frac{i}{(2\pi)^{3}}\int \epsilon({\tilde p})
\delta({\tilde p}^{2}-\mu^{2})\ e^{-i {\tilde p}\cdot
{x}}\,\mathrm{d}^{4}{\tilde p} , \end{equation} with
$\epsilon({\tilde p})=1$ for $\omega > 0$ and $\epsilon({\tilde
p})=-1$ for $\omega < 0$. Since $\Delta({x},\,\mu)$ vanishes for
space-like distances $x^2 < 0$ \cite{xx} we find that microcausality
is fulfilled also in the interacting case  (cf. Ref. \cite{Peter}).
Since the DQPM - as an effective approach to QCD - employs spectral
functions  of the type (\ref{eq:TheorieErgebnis}) (or (\ref{GV})
and (\ref{GF})) we may conclude that the model approach conserves
microcausality strictly \footnote{ Note also that the spectral
function (\ref{eq1}) corresponds to a Breit-Wigner representation of
the integrand $\epsilon({\tilde p}) \delta({\tilde p}^{2}-\mu^{2})$
in (\ref{Delta}) for a finite width $\gamma$.}.
\begin{figure}[hbt]
   \centering
   \includegraphics[width=0.95\linewidth]{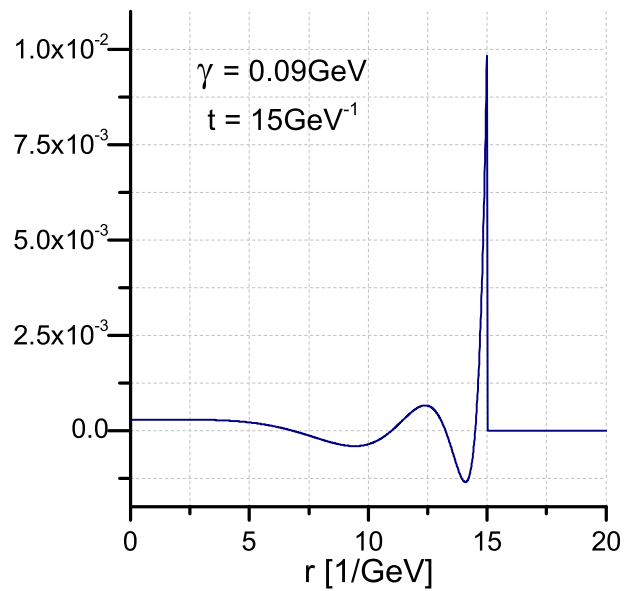}
   \caption{(Color online) The regular part of the Green function (\ref{regular})
   at time $t=15$/GeV as a function of
   the distance $r$. The analytical and numerical results from the integration on the grid
   are identical within the linewidth. }
   \label{fig1}
\end{figure}

\section{Space-like and time-like momentum contributions}
We now come to the central question of our study: Is microcausality
fulfilled in the four-momentum integral \begin{equation}
\label{timel} C(x)=\int \frac{d\omega}{2 \pi} \frac{d^3 p}{(2
\pi)^3} \ \Im(G_{ret}(\omega,{\bf p}))\ \exp(-i(\omega t -  {\bf p}
\cdot \boldsymbol{x}))  \end{equation} when restricting to time-like
$\Theta(\omega^2-{\bf p}^2)$ or space-like $\Theta({\bf
p}^2-\omega^2)$ four-momenta?

To answer this question we can no longer perform the contour
integration over $d\omega$ due to the $\Theta$-functions in
four-momentum and have to evaluate the integrals (\ref{timel})
numerically exploiting the antisymmetry of the integrand (\ref{eq1})
in $\omega$ and carrying out the angular integration in the
three-momentum. This leads to
\begin{equation} \label{int1}
C({x})=  -\frac{i \gamma}{2 \pi^3 \left|\boldsymbol{x}\right|}
\intop_{0}^{\infty} dp \intop_{0}^{\infty} d\omega \  \sin(\omega t)
\sin(\left|\boldsymbol{x}\right|p)
\end{equation} $$ \times
\frac{ p}{{\tilde E}} \biggl(\frac{1}{(\omega - \tilde{E})^2 +
\gamma^2} - \frac{1}{(\omega + \tilde{E})^2 + \gamma^2} \biggr) \,$$
using (\ref{eq1})  which can be 'solved' on a numerical grid as well
as by analytical integration (cf. Section II). As mentioned before
the integral (\ref{int1}) has a singular part ($\delta(t-r)/(4\pi
r)$ using $r=\left|\boldsymbol{x}\right|$) as well as a regular part
given by (\ref{regular}). The singular part can be subtracted in the
integral (\ref{int1}) - to achieve a better convergence - by
considering
\begin{equation} \label{int2} C({x})-
\frac{\delta(t-r)}{4\pi r} e^{- \gamma t}
=  -\frac{i}{2 \pi^3 r} \intop_{0}^{\infty} dp \intop_{0}^{\infty}
d\omega \ \sin(\omega t) \sin(r p)
\end{equation} $$ \times
\frac{8 p \omega (M^2-\gamma^2) \left[ \omega^2 - p^2 -
(\gamma^2+M^2)/2 \right]} {\left[(\omega^2-p^2-M^2)^2+4\omega^2
\gamma^2 \right] \left[(\omega^2-p^2-\gamma^2)^2+4\omega^2 \gamma^2
\right]} . $$
 In this way the $\delta$-distribution on the light-cone (decaying exponentially in time) is
subtracted explicitly on the same computational grid. In order to
demonstrate the validity of this numerical subtraction scheme we
show in Fig. \ref{fig1} a comparison of the analytical result
(\ref{regular}) with the corresponding numerical evaluation of
(\ref{int2}) for $M$ = 1 GeV and $\gamma=0.3$ GeV at $t=15$/GeV.
Indeed, both results agree within the linewidth and are identical to
zero for $r > t$.

\begin{figure}[hbt]
   \centering
   \includegraphics[width=0.9\linewidth]{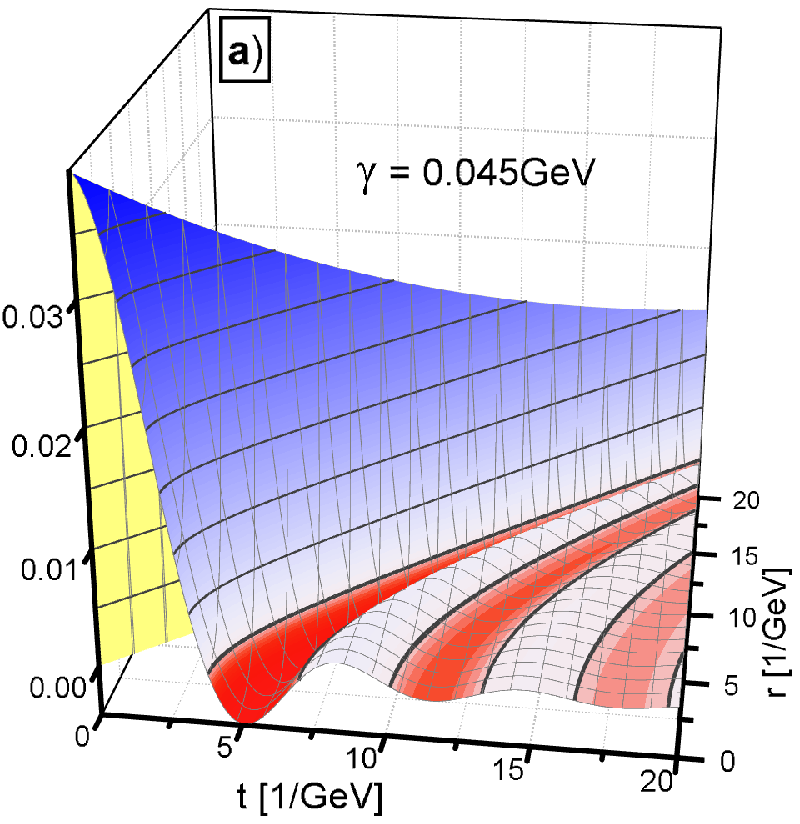}
   \includegraphics[width=0.9\linewidth]{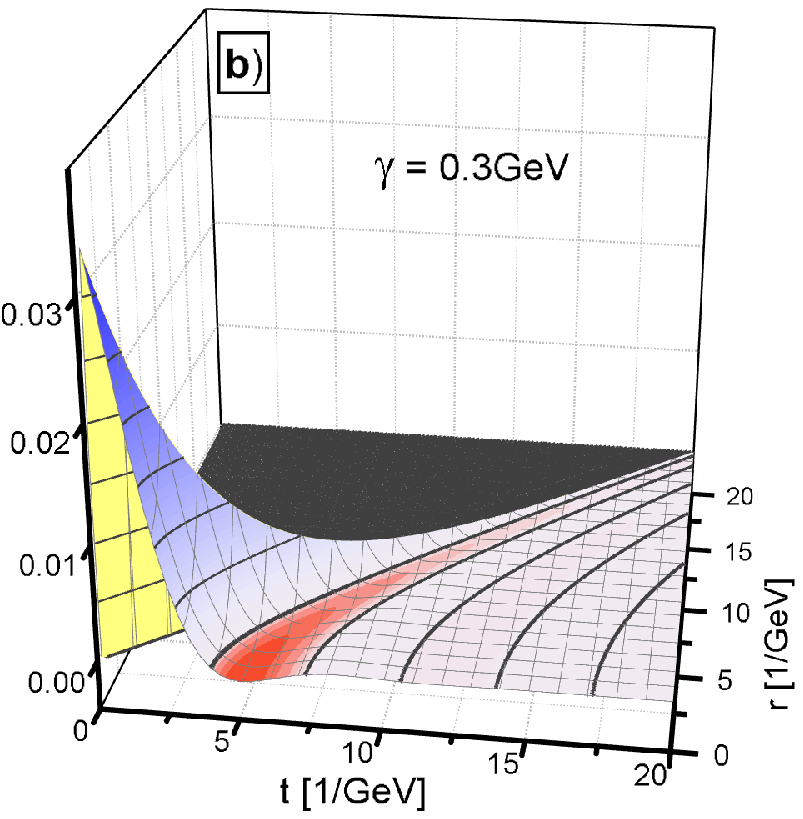}
   \caption{(Color online) The regular part of the commutator (\ref{regular})  as a function of
   the distance $r$ and time $t$ for $\gamma$=0.045 GeV (a) and $\gamma$ = 0.3 GeV (b).}
   \label{fig2}
\end{figure}

In Fig. \ref{fig2} we show the regular part
(\ref{regular})   for a width $\gamma$=0.045 GeV
(a) and $\gamma$ = 0.3 GeV (b). The signal decays exponentially in
time ($\sim \exp(-\gamma t)$)  and shows hyperbolic
oscillations within the lightcone while being zero outside the lightcone. The
numerical results and the analytical expression (\ref{regular}) are
identical on the level of three digits for both cases.

\begin{figure}[hbt]
   \centering
   \includegraphics[width=0.9\linewidth]{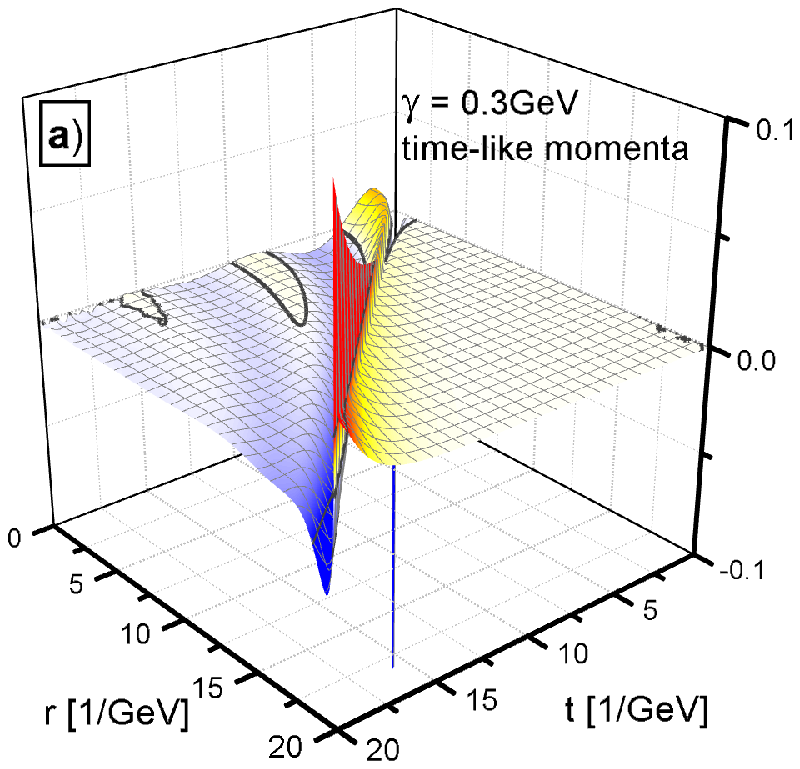}
   \includegraphics[width=0.9\linewidth]{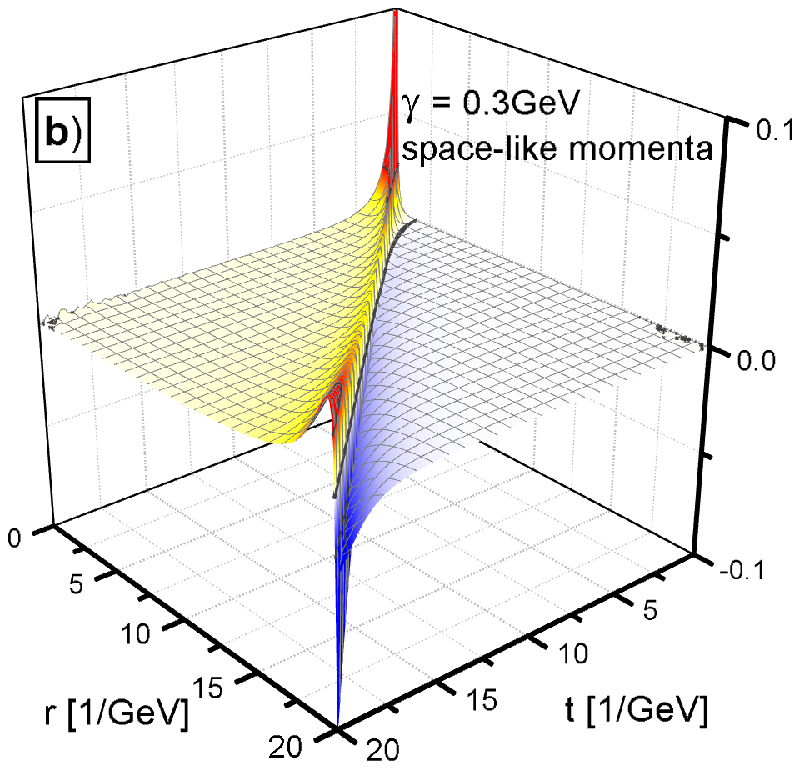}
   \caption{(Color online) The regular part of the commutator (\ref{regular}) (multiplied by $\exp( \gamma t)$)
    as a function of
   the distance $r$ and time $t$ for $\gamma$=0.3 GeV for time-like four-momenta (a) and
   space-like four-momenta (b).}
   \label{fig2b}
\end{figure}

We now turn to the numerical results for strong coupling when gating
on time-like and space-like four momenta in (\ref{int2}) separately.
The results are displayed in Fig. \ref{fig2b} for $\gamma$ = 0.3 GeV
when including only time-like momenta (a) or only space-like momenta
(b). Note that the numerical results in Fig. \ref{fig2b} have been
multiplied by $\exp( \gamma t)$ in order to compensate for the
exponential decay in time.

\begin{figure}[h]
   \centering
   \includegraphics[width=0.95\linewidth]{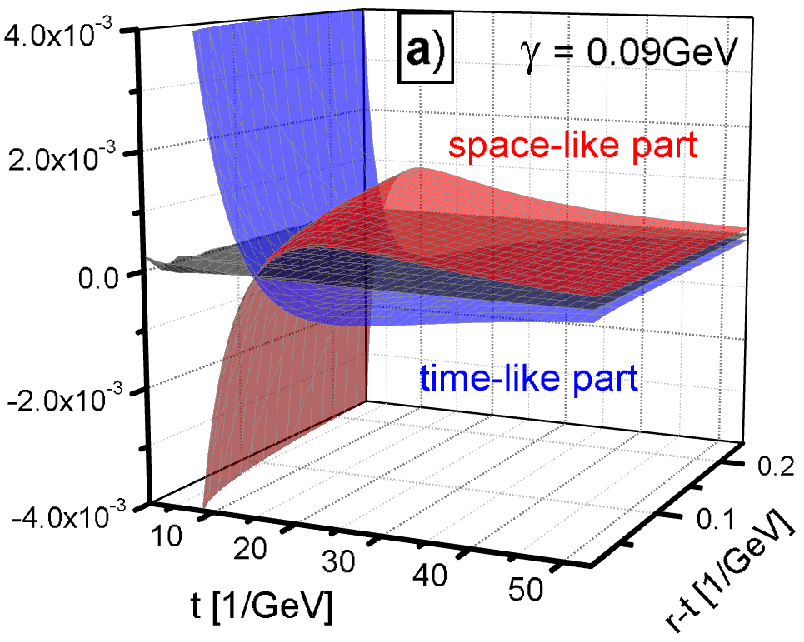}
   \includegraphics[width=0.95\linewidth]{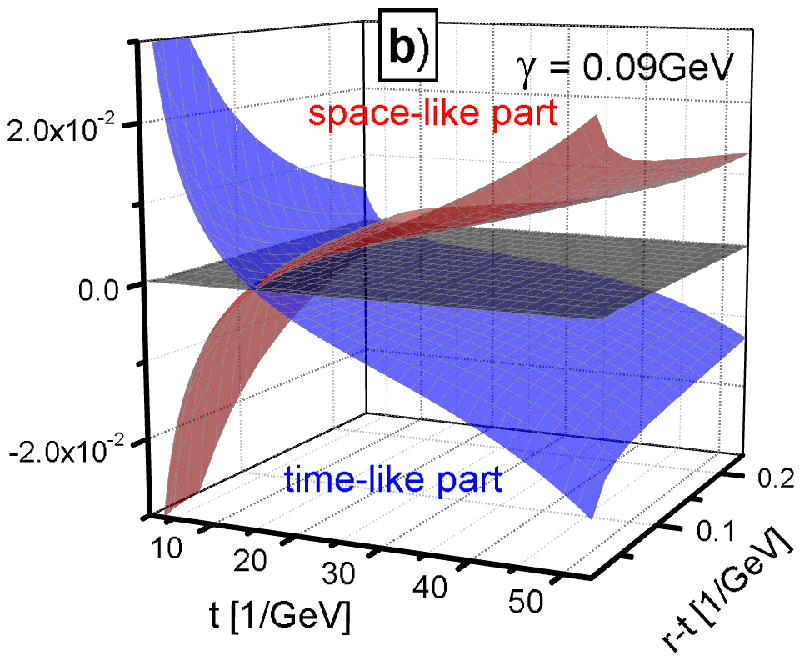}
   \caption{(Color online) (a) The regular part of the commutator (\ref{regular})  as a function of
   the  time $t$ and $r-t$ in the space-like region for $\gamma$=0.09 GeV.
   The contribution from space-like
   momenta is given by the  red area while the contribution from time-like
   momenta is given by the  blue area which is of opposite sign. (b) Same as Fig. (a) but multiplied by
   $\exp(\gamma t)$.}
   \label{figg1}
\end{figure}

It is seen that both results do not vanish for $r > t$ and thus
violate microcausality. This is shown more explicitly in Fig.
\ref{figg1} (a) as a function of time $t$ and $r-t$ in the space-like
region and demonstrates that both contributions are nonvanishing but
of opposite sign such that their sum becomes identically zero. In Fig.
\ref{figg1} (b) we display the same quantities as in (a) but multiplied by $\exp(\gamma t)$
which demonstrates that both contributions do not decay exponentially in time as
seen from the full analytical solution (\ref{eq:TheorieErgebnis}).
This clearly demonstrates that a restriction to either space-like or
time-like momentum parts of the spectral function violates causality
while both parts together conserve microcausality in line with the
analytical result in Section II. The violation of microcausality is
tiny in case of $\gamma \ll M$ but becomes sizeable for $\gamma >$
0.1 GeV.

\begin{figure}[hbt]
   \centering
   \includegraphics[width=0.95\linewidth]{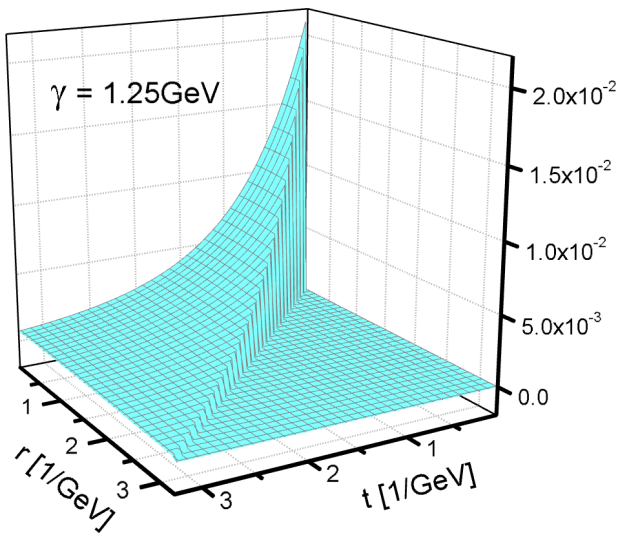}
   \caption{(Color online) The regular part of the commutator (\ref{regular})  as a function of
   the distance $r$ and time $t$ for $\gamma$=1.25 GeV.}
   \label{fig7}
\end{figure}

When considering the 'aperiodic' limit $\gamma \rightarrow M$, i.e.
$\mu \rightarrow 0$, we find from (\ref{regular1}) that ${\tilde R}(t^2-r^2)$
vanishes identically for $\mu =0$ and the commutator
(\ref{eq:ZusammenhangKommutatorGreen}) only has support on the
lightcone. Furthermore, in the limit $\mu \rightarrow 0$ the
oscillations in $R(t^2-r^2)$ vanish as can be extracted from a
Taylor expansion of (\ref{regular}) (with $J_1(z) \approx z/2 \pm
\cdots$)
\begin{eqnarray}
&& R(t^{2}-r^{2})\approx \Theta\left(t^{2}-r^{2}\right)
\frac{\mu}{8\pi\sqrt{t^{2}-r^{2}}}
(\mu\sqrt{t^{2}-r^{2}}) \cdots   \nonumber\\
&& =\Theta\left(t^{2}-r^{2}\right)
\frac{\mu^2}{8\pi} \cdots  .\label{Taylor}
\end{eqnarray}
For $\gamma > M$ (overdamped fields) we no longer find oscillations
of the regular part (\ref{regular}) within the lightcone but only an
exponentially decaying signal as seen from Fig. \ref{fig7} for
$\gamma =$ 1.25 GeV.

We close in pointing out that our numerical scheme allows to employ
almost arbitrary spectral functions in (\ref{int1}) and to check
if microcausality holds. Without explicit representation we note
that using a three-momentum width $\gamma({\bf p}^2)$ in the
spectral function (\ref{eq1}) the commutator (\ref{int1}) no
longer vanishes for $r > t$ since the individual momentum modes
decay on different time scales and the field equation
(\ref{eq:KleinGordon}) becomes non-local in this case. Nevertheless,
the normalization condition (\ref{equ:Sec2.3}) still is fulfilled.  Furthermore,
the commonly adopted form,
\begin{equation}
\label{eq11}
 A (\omega, {\bf p})= \frac{2 M \gamma}{(\omega^2
- {\bf p}^2 - M^2)^2 + 4 \gamma^2 M^2}, \end{equation}
 also violates microcausality.

\section{Summary}
In this study we have examined effective propagators of the type
(\ref{eq:G(p)AlgebraischGewonnen}) as used e.g. in the Dynamical
QuasiParticle Model (DQPM) \cite{Peshier,DQPM1} for an approximation
to QCD propagators at temperatures above the critical temperature
$T_c$ for deconfinement. It could be shown analytically that their
spectral functions (or imaginary parts) do not lead to a violation
of microcausality, i.e. to a vanishing commutator of the interacting
fields outside the lightcone. However, when restricting to only
space-like or time-like four-momentum contributions of the spectral
function a violation of microcausality is found numerically which
becomes severe in case of strong coupling. Moreover, the
space-like or time-like four-momentum contributions separately no
longer decay exponentially in time as in case of the full solution
 (\ref{eq:TheorieErgebnis}). Furthermore, we have
found that using a three-momentum dependent width $\gamma({\bf p}^2)$ in the
spectral function (\ref{eq1}) the commutator (\ref{int1}) no longer
vanishes for $r > t$ since the individual momentum modes decay on
different time scales and the field equation (\ref{eq:KleinGordon})
becomes non-local in space  in this case.
This also holds for the spectral function (\ref{eq11})
which is often employed in phenomenological models.

Our findings imply  that the
modeling of effective field theories for strongly interacting
systems has to be considered with great care and restrictions to
time-like four momenta in case of broad spectral functions have to
be ruled out.
\\

The authors acknowledge valuable discussions with B.-J. Schaefer, L.
von Smekal and T. Steinert.
\vspace{-0.5cm}

\end{document}